\documentclass{PoS}

\def\Slash#1{\not\!\!#1}

\title{Analytical Relation between Quark Confinement and 
Chiral Symmetry Breaking in QCD}

\ShortTitle{Analytical relation between confinement and chiral symmetry breaking}

\ShortTitle{Analytical relation between Polyakov loop and
Dirac eigenvalues in odd-$N_t$ lattice QCD}

\author{\speaker{Hideo Suganuma}, Takahiro M. Doi \\
        Department of Physics \& Division of Physics and Astronomy, 
Graduate School of Science, \\
Kyoto University, 
Kitashirakawaoiwake, Sakyo, Kyoto 606-8502, Japan\\
        E-mail: \email{suganuma@scphys.kyoto-u.ac.jp}}

\author{Takumi Iritani \\
High Energy Accelerator Research Organization (KEK), 
Tsukuba, Ibaraki 305-0801, Japan}

\abstract{
We study the relation between quark confinement and 
spontaneous chiral-symmetry breaking directly in QCD. 
In lattice QCD formalism, 
we derive an analytical gauge-invariant relation between the Polyakov loop 
$\langle L_P \rangle$ and the Dirac eigenvalues $\lambda_n$, i.e., 
$\langle L_P \rangle \propto 
\sum_n \lambda_n^{N_t -1} \langle n|\hat U_4|n \rangle$, 
on a temporally odd-number lattice, where the temporal lattice size $N_t$ 
is odd. 
Here, $|n \rangle$ denotes the Dirac eigenmode, i.e., 
$\Slash D|n \rangle=i\lambda_n|n \rangle$, and 
$\hat U_4$ the temporal link-variable operator.
We here use an ordinary square lattice with 
the normal periodic boundary condition 
for link-variables $U_\mu(s)$ in the temporal direction. 
Because of the factor $\lambda_n^{N_t -1}$ in the analytical relation, 
the contribution of low-lying Dirac modes to the Polyakov loop is 
negligibly small in both confined and deconfined phases, 
while the low-lying Dirac modes are essential for chiral symmetry breaking. 
Also, in lattice QCD simulations, 
we numerically confirm the analytical relation, 
non-zero finiteness of $\langle n|\hat U_4|n \rangle$ for each Dirac mode, 
and negligibly small contribution of low-lying Dirac modes 
to the Polyakov loop. 
Thus, we conclude that low-lying Dirac modes 
are not essential for confinement, 
which indicates no direct one-to-one correspondence between 
confinement and chiral symmetry breaking in QCD.
}

\FullConference{XV International Conference on Hadron Spectroscopy \\
                 4-8/11/2013\\
                 Nara, Japan}

\begin{document}

\section{Introduction: Are color confinement and CSB one-to-one in QCD?}

QCD has two outstanding 
nonperturbative phenomena of color confinement and 
spontaneous chiral-symmetry breaking (CSB) \cite{NJL61}.
However, their relation is not yet known directly from QCD, 
and to clarify their precise relation is one of the important problems 
in theoretical physics \cite{SST95,M95W95,GIS12}. 

For quark confinement, 
the Polyakov loop $\langle L_P \rangle$ is 
one of the typical order parameters, and 
it relates to the single-quark free energy $E_q$ as 
$\langle L_P \rangle \propto e^{-E_q/T}$ 
at temperature $T$. 
Also, its fluctuation is recently found to be important 
in the QCD phase transition \cite{LFKRS13}.
For CSB, the order parameter 
is the chiral condensate $\langle \bar qq \rangle$,
and low-lying Dirac modes play the essential role \cite{BC80}.

A strong correlation between confinement and CSB has been suggested by 
almost coincidence between deconfinement and chiral-restoration temperatures 
\cite{Rothe12}, although slight difference of about 25MeV between them 
is pointed out in recent lattice QCD studies \cite{AFKS06}.
Their correlation has been also suggested 
in terms of QCD-monopoles \cite{SST95,M95W95}, 
which topologically appear in QCD in the maximally Abelian gauge. 
By removing the monopoles from the QCD vacuum, 
confinement and CSB are simultaneously lost in lattice QCD 
\cite{SST95,M95W95}. (See Fig.1.)
This indicates an important role of QCD-monopoles 
to both confinement and CSB, and thus 
these two phenomena seem to be related via the monopole.
However, the direct relation of confinement and CSB 
is still unclear.

Then, we have a question. {\it if only the relevant ingredient of 
CSB is carefully removed from the QCD vacuum, 
how will be quark confinement?}
In this study, we derive an analytical relation between 
the Polyakov loop and the Dirac modes in temporally odd-number lattice QCD, 
where the temporal lattice size is odd, 
and discuss the relation between confinement and CSB \cite{SDI13,DSI13}.

\begin{figure}[h]
\begin{center}
\includegraphics[scale=0.32]{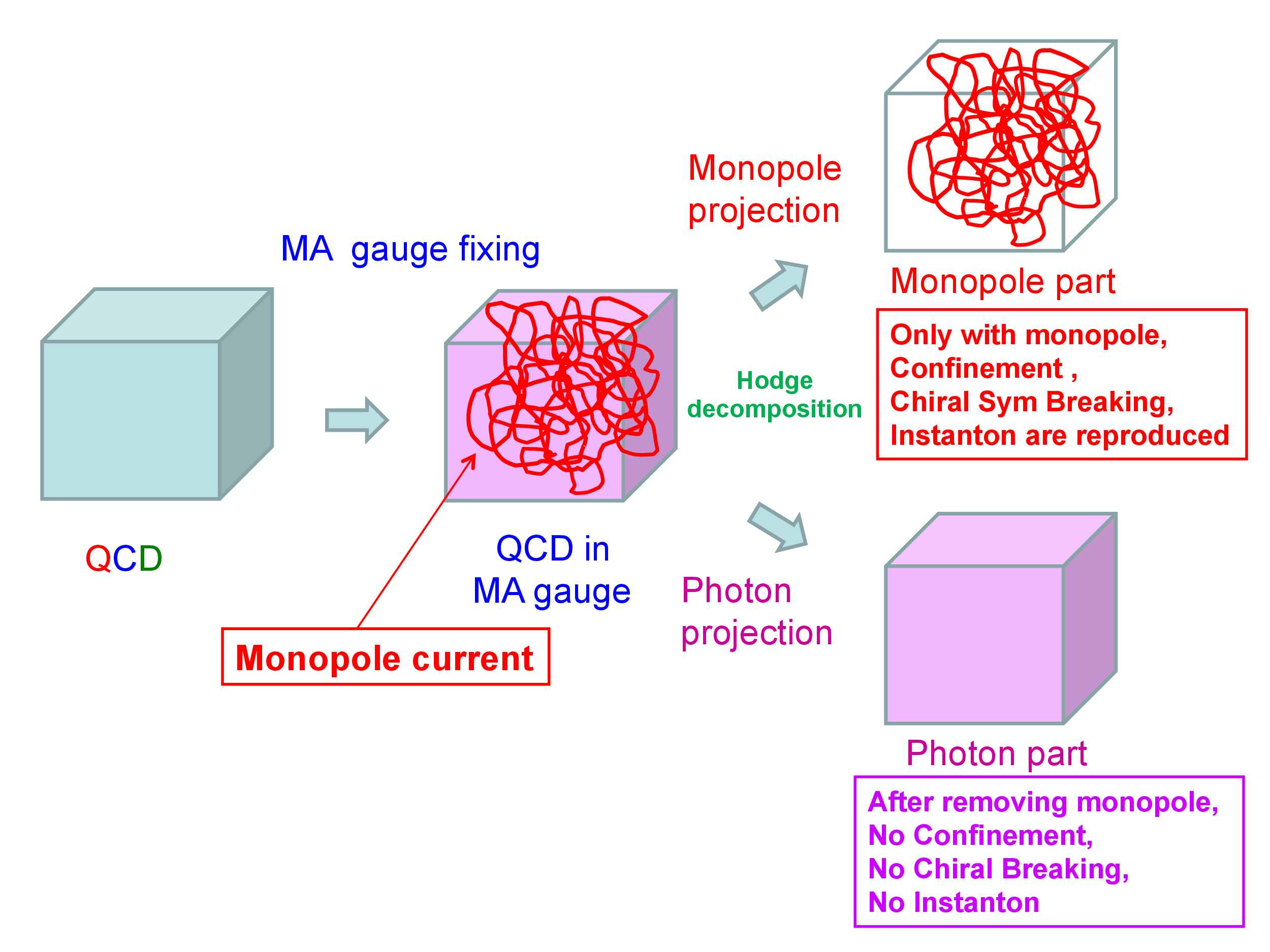}
\hspace{0.5cm} \includegraphics[scale=0.3]{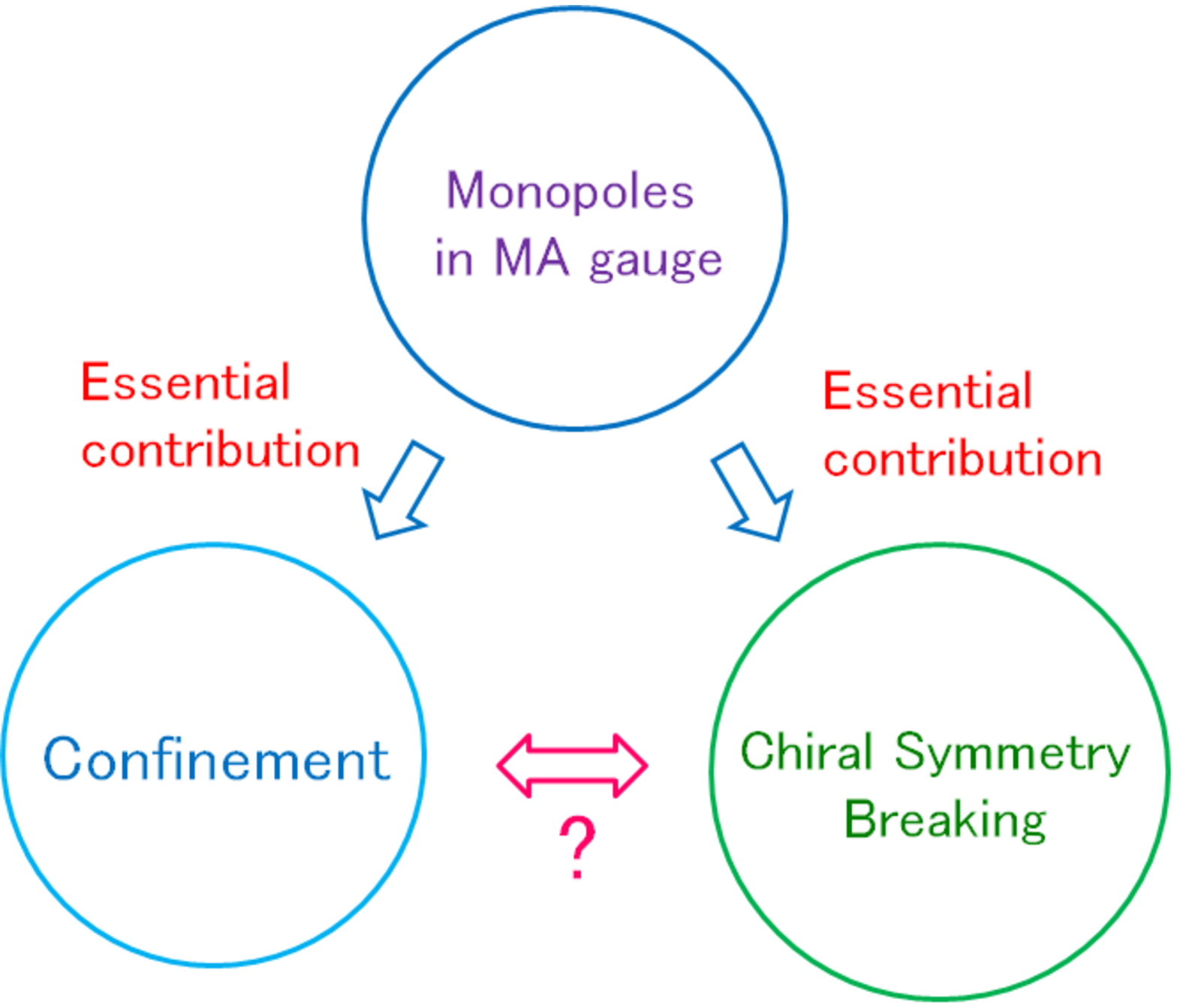}
\vspace{-0.1cm}
\caption{
In the MA gauge, QCD-monopoles topologically appear. 
By removing the monopole from the QCD vacuum, 
confinement and CSB are simultaneously lost \cite{SST95,M95W95}. 
This means essential role 
of monopoles to both confinement and CSB. 
However, the direct relation between confinement and CSB is unclear.
}
\end{center}
\vspace{-1cm}
\end{figure}

\section{Lattice QCD formalism for 
Dirac operator, Dirac eigenvalues and Dirac modes}

Note that, in our studies \cite{SDI13,DSI13}, we just consider 
the mathematical expansion by eigenmodes $|n \rangle$ of 
the Dirac operator $\Slash D=\gamma_\mu D_\mu$, 
using the completeness of $\sum_n|n \rangle \langle n|=1$. 
In general, instead of $\Slash D$, 
one can consider any (anti)hermitian operator, e.g., $D^2=D_\mu D_\mu$, 
and the expansion in terms of its eigenmodes. 
In this paper, to consider CSB, 
we adopt $\Slash D$ and the expansion by its eigenmodes.

We use an ordinary square lattice with spacing $a$ and 
size $V \equiv N_s^3 \times N_t$, and impose 
the normal periodic boundary condition 
for the link-variable $U_\mu(s)={\rm e}^{iagA_\mu(s)}$ 
($A_\mu$: gluon fields) in the temporal direction. 
In lattice QCD, the Dirac operator 
$\Slash D = \gamma_\mu D_\mu$ is expressed with 
$U_\mu(s)$ as
\begin{eqnarray}
 \Slash{D}_{s,s'} 
 \equiv \frac{1}{2a} \sum_{\mu=1}^4 \gamma_\mu 
\left[ U_\mu(s) \delta_{s+\hat{\mu},s'}
 - U_{-\mu}(s) \delta_{s-\hat{\mu},s'} \right],
\end{eqnarray}
where $\hat\mu$ is the unit vector in $\mu$-direction in the lattice unit, 
and $U_{-\mu}(s)\equiv U^\dagger_\mu(s-\hat \mu)$.
Adopting hermitian $\gamma$-matrices as $\gamma_\mu^\dagger=\gamma_\mu$, 
the Dirac operator $\Slash D$ is anti-hermitian and satisfies 
$\Slash D_{s',s}^\dagger=-\Slash D_{s,s'}$.
We introduce 
the normalized Dirac eigen-state $|n \rangle$ 
and the Dirac eigenvalue $i\lambda_n$ ($\lambda_n \in {\bf R}$),
\vspace{-0.1cm}
\begin{eqnarray}
\Slash D |n\rangle =i\lambda_n |n \rangle, \qquad
\langle m|n\rangle=\delta_{mn}, \qquad \sum_n |n \rangle \langle n|=1.
\end{eqnarray}
The Dirac eigenfunction $\psi_n(s)\equiv\langle s|n \rangle$ 
satisfies 
$\sum_{s'}\Slash D_{s,s'} \psi_n(s')=i\lambda_n \psi_n(s)$ 
and gauge-transforms as 
$\psi_n(s)\rightarrow V(s) \psi_n(s)$,
which is the same as the quark field, 
apart from an irrelevant global phase \cite{GIS12}.

Now, we define the link-variable operator $\hat U_{\pm \mu}$ \cite{GIS12}
by the matrix element of 
\vspace{-0.15cm}
\begin{eqnarray}
\langle s |\hat U_{\pm \mu}|s' \rangle 
=U_{\pm \mu}(s)\delta_{s\pm \hat \mu,s'}.
\end{eqnarray}
With the link-variable operator, 
the Dirac operator and covariant derivative 
are simply expressed, 
\vspace{-0.15cm}
\begin{eqnarray}
\Slash{\hat D}
=\frac{1}{2a}\sum_{\mu=1}^{4} \gamma_\mu (\hat U_\mu-\hat U_{-\mu}),
\qquad 
\hat D_\mu=\frac{1}{2a}(\hat U_\mu-\hat U_{-\mu}).
\label{eq:Dop}
\end{eqnarray}
The Polyakov loop is also simply written as the functional trace of 
$\hat U_4^{N_t}$, i.e.,
$
\langle L_P \rangle
=\frac{1}{N_c V} \langle {\rm Tr}_c \{\hat U_4^{N_t}\}\rangle,
$
where, ``${\rm Tr}_c$'' denotes the functional trace 
of ${\rm Tr}_c \equiv \sum_s {\rm tr}_c$ with 
the trace ${\rm tr}_c$ over color index.

\section{Direct relation between Polyakov loop and Dirac modes 
on odd-number lattice}

Now, we consider a temporally odd-number lattice \cite{SDI13,DSI13}, 
where the temporal lattice size $N_t$ is odd. 
The spatial lattice size $N_s$ is taken to be larger than $N_t$, 
i.e., $N_s > N_t$. 
Note that, in the continuum limit of $a \rightarrow 0$ and 
$N_t \rightarrow \infty$, 
any number of large $N_t$ gives the same physical result.
Then, it is no problem to use the odd-number lattice.

In general, only gauge-invariant quantities 
such as closed loops and the Polyakov loop 
survive in QCD, according to the Elitzur theorem \cite{Rothe12}.
All the non-closed lines are gauge-variant 
and their expectation values are zero.
Note here that any closed loop 
needs even-number link-variables on the square lattice, 
except for the Polyakov loop.
 (See Fig.2.)
On the temporally odd-number lattice, 
we consider the following functional trace 
and its expectation value \cite{SDI13,DSI13}: 
\vspace{-0.15cm}
\begin{eqnarray}
I\equiv {\rm Tr}_{c,\gamma} (\hat{U}_4\hat{\Slash{D}}^{N_t-1}), \qquad
 \langle I\rangle=\langle {\rm Tr}_{c,\gamma} (\hat{U}_4\hat{\Slash{D}}^{N_t-1})\rangle.
\label{eq:FTV}
\end{eqnarray}
Here, 
${\rm Tr}_{c,\gamma}\equiv \sum_s {\rm tr}_c 
{\rm tr}_\gamma$ includes 
${\rm tr}_c$ 
and the trace ${\rm tr}_\gamma$ over spinor index.
The expectation value $\langle I \rangle$ 
is obtained as the gauge-configuration average in lattice QCD.
In the case of enough large volume $V$, one can expect 
$\langle O \rangle \simeq {\rm Tr}~O/{\rm Tr}~1$ 
for any operator $O$ at each gauge configuration.

\begin{figure}[h]
\begin{center}
\includegraphics[scale=0.25]{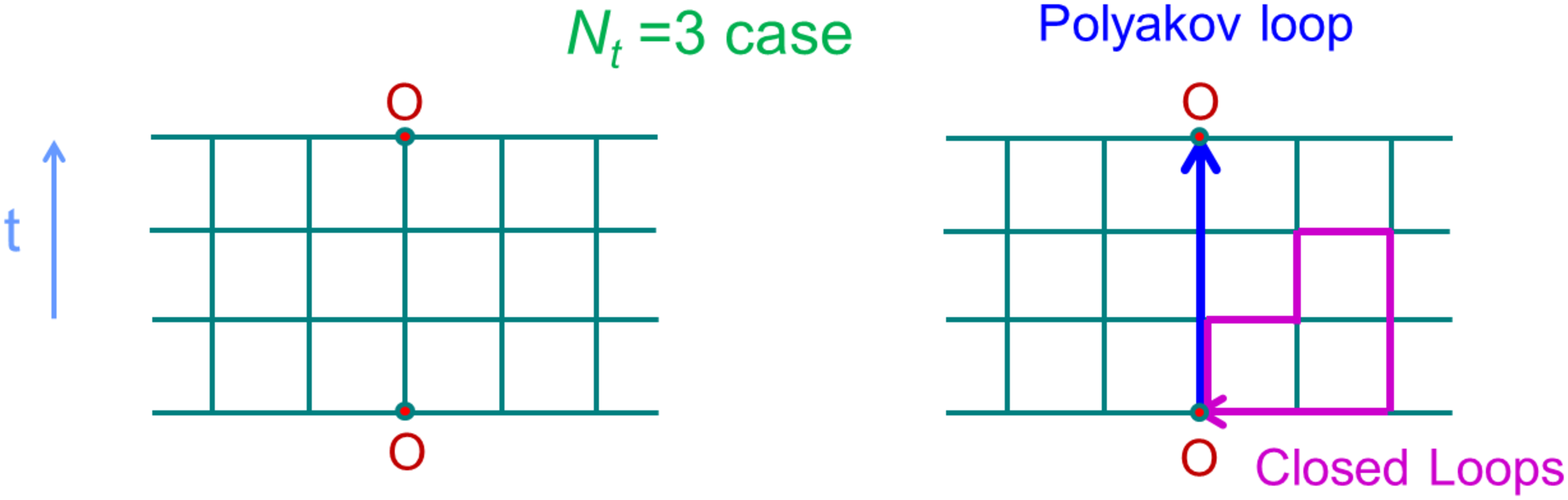}
\caption{
An example 
%($N_t=3$ case) 
of the temporally odd-number lattice.
Only gauge-invariant quantities such as 
closed loops and the Polyakov loop survive in QCD.
Closed loops have even-number links on the square lattice.
}
\end{center}
\vspace{-0.3cm}
\end{figure}

From Eq.(\ref{eq:Dop}), 
$\hat U_4\Slash{\hat D}^{N_t-1}$ 
is expressed as a sum of products of $N_t$ link-variable operators, 
since the Dirac operator $\Slash{\hat D}$ 
includes one link-variable operator in each direction of $\pm \mu$.
Then, $\hat U_4\Slash{\hat D}^{N_t-1}$ 
includes many trajectories with the total length $N_t$ 
(in the lattice unit) 
on the square lattice, as shown in Fig.3.
Note that all the trajectories with the odd-number length $N_t$ 
cannot form a closed loop 
on the square lattice, and thus give gauge-variant contribution, 
except for the Polyakov loop.

\begin{figure}[h]
\begin{center}
\includegraphics[scale=0.23]{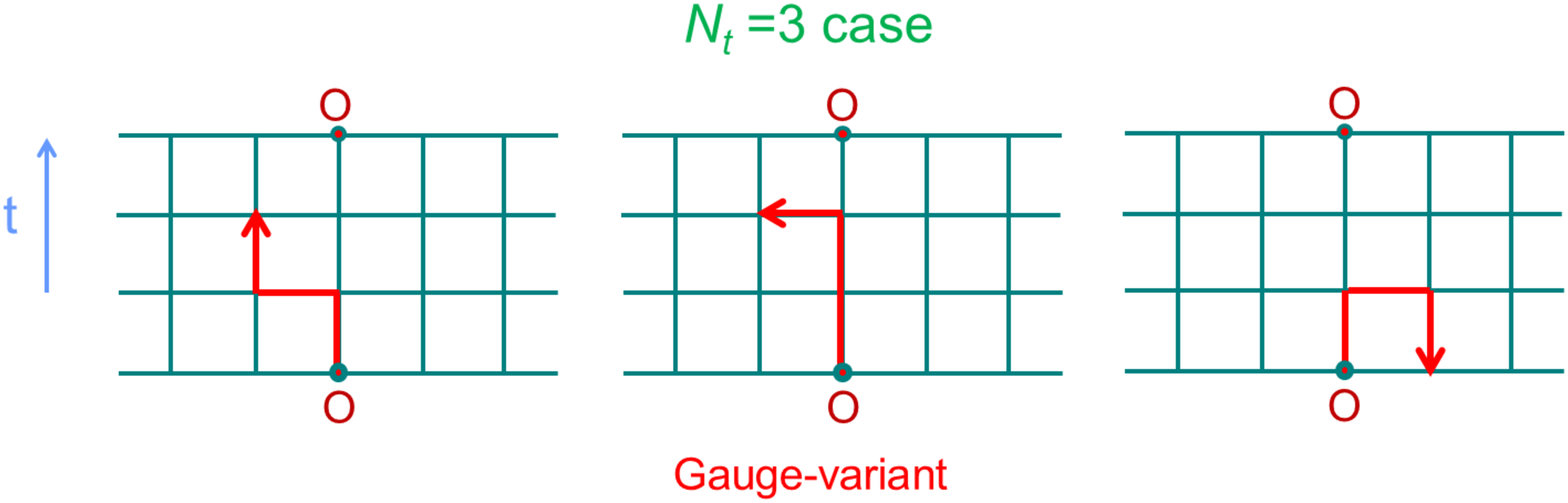}
\hspace{0.4cm}
\includegraphics[scale=0.23]{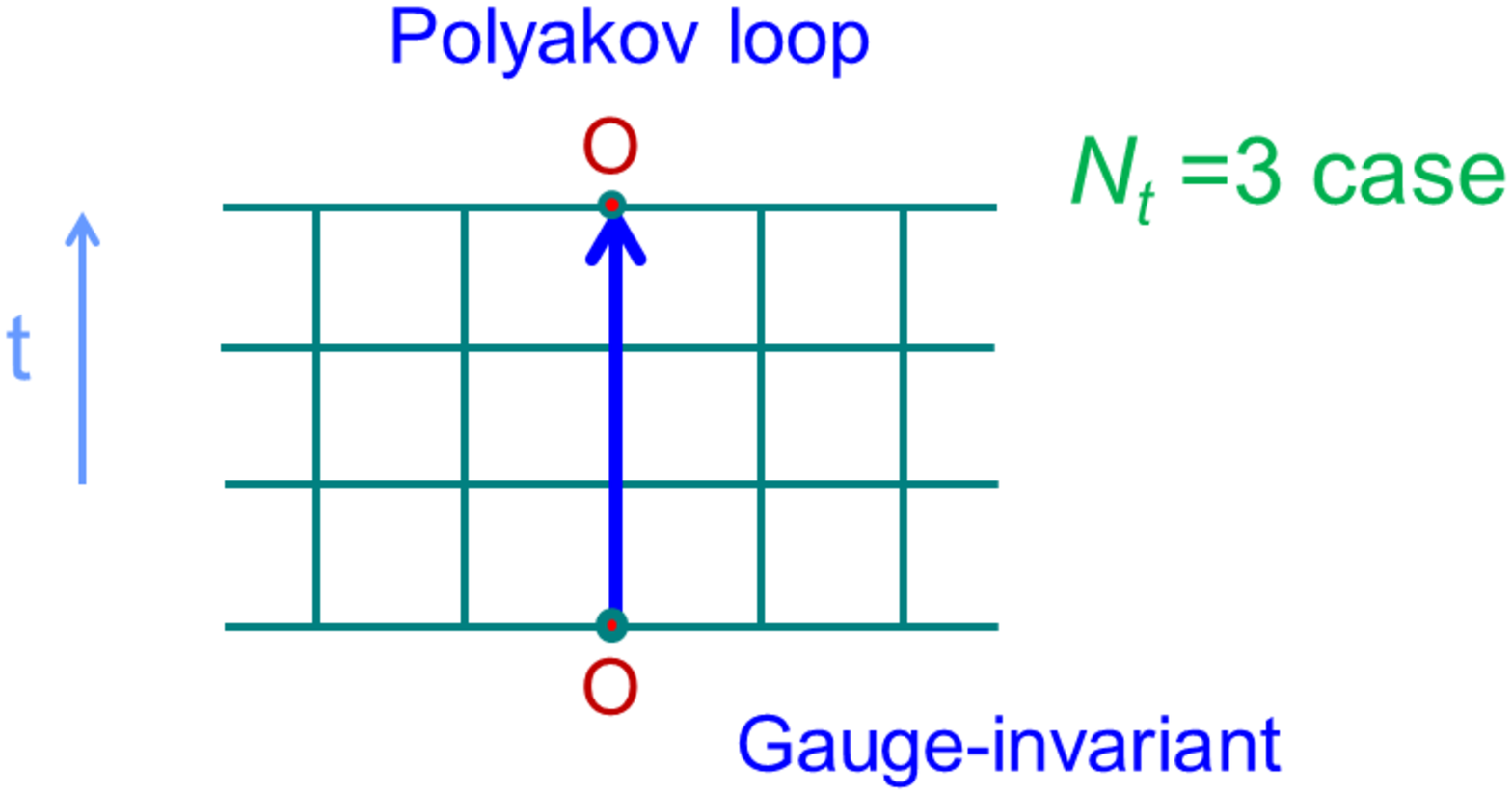}
\vspace{-0.7cm}
\caption{
Examples of the trajectories stemming from 
$\langle I \rangle =
\langle {\rm Tr}_{c,\gamma}(\hat U_4\Slash{\hat D}^{N_t-1})\rangle$. 
For each trajectory, the total length is 
$N_t$, and the ``first step'' is positive 
temporal direction corresponding to $\hat U_4$.
All the trajectories with the odd-number length $N_t$ 
cannot form a closed loop on the square lattice,
so that they are gauge-variant and give no contribution, 
except for the Polyakov loop.
Thus, only the Polyakov-loop ingredient survives in $\langle I \rangle$. 
}
\end{center}
\vspace{-0.3cm}
\end{figure}

Hence, among the trajectories stemming from 
$\langle I \rangle
=\langle {\rm Tr}_{c,\gamma}(\hat U_4\Slash{\hat D}^{N_t-1}) \rangle$, 
all the non-loop trajectories are gauge-variant and give no contribution, 
according to the Elitzur theorem \cite{Rothe12}.
Only the exception is the Polyakov loop. (See Fig.3.)
Thus, in $\langle I \rangle$, 
only the Polyakov-loop ingredient can survive 
as the gauge-invariant quantity, and 
$\langle I \rangle$ is proportional to the Polyakov loop $\langle L_P \rangle$.

Actually, we can mathematically derive the following relation \cite{SDI13}:
\begin{eqnarray}
\langle I\rangle
&=&\langle {\rm Tr}_{c,\gamma} (\hat U_4 \hat{\Slash D}^{N_t-1}) \rangle
=\langle {\rm Tr}_{c,\gamma} \{\hat U_4 (\gamma_4 \hat D_4)^{N_t-1}\} \rangle
~~{\rm 
(
\raisebox{1.2ex}{.}\raisebox{.2ex}{.}\raisebox{1.2ex}{.} 
~only~gauge\hbox{-}invariant~terms~survive)} 
\nonumber \\
&=&4\langle {\rm Tr}_{c} (\hat U_4 \hat D_4^{N_t-1}) \rangle
\qquad \quad \qquad \qquad \qquad \qquad (
\raisebox{1.2ex}{.}\raisebox{.2ex}{.}\raisebox{1.2ex}{.} 
~\gamma_4^{N_t-1}={1}, 
~{\rm tr}_\gamma {1}=4) 
\nonumber \\
&=&\frac{4}{(2a)^{N_t-1}}
\langle {\rm Tr}_{c} \{\hat U_4 (\hat U_4-\hat U_{-4})^{N_t-1}\} \rangle
\quad \qquad ~~~(
\raisebox{1.2ex}{.}\raisebox{.2ex}{.}\raisebox{1.2ex}{.} 
~\hat D_4=\frac{1}{2a}(\hat U_4-\hat U_{-4}))
\nonumber \\
&=&\frac{4}{(2a)^{N_t-1}} \langle {\rm Tr}_{c} \{ \hat U_4^{N_t} \}\rangle
=\frac{12V}{(2a)^{N_t-1}}\langle L_P \rangle. 
\qquad ~~{\rm 
(\raisebox{1.2ex}{.}\raisebox{.2ex}{.}\raisebox{1.2ex}{.} 
~only~gauge\hbox{-}invariant~terms~survive)}~~~
\label{eq:FTtoPL}
\end{eqnarray}

On the other hand, we calculate the functional trace 
in Eq.(\ref{eq:FTV}) using the complete set of 
the Dirac-mode basis $|n\rangle$ satisfying $\sum_n |n\rangle \langle n|=1$, 
and find the Dirac-mode representation of 
\begin{eqnarray}
 \langle I\rangle=\sum_n\langle n|\hat{U}_4\Slash{\hat{D}}^{N_t-1}|n\rangle
=i^{N_t-1}\sum_n\lambda_n^{N_t-1}\langle n|\hat{U}_4| n \rangle. 
\label{eq:FTtoD}
\end{eqnarray}
Combing Eqs.(\ref{eq:FTtoPL}) and (\ref{eq:FTtoD}), we obtain the analytical 
relation between $ \langle L_P \rangle$ and $\lambda_n$ 
in QCD \cite{SDI13,DSI13}: 
\begin{eqnarray}
\langle L_P \rangle=\frac{(2ai)^{N_t-1}}{12V}
\sum_n\lambda_n^{N_t-1}\langle n|\hat{U}_4| n \rangle. 
\label{eq:PLvsD}
\end{eqnarray}
This is a direct relation between the Polyakov loop $\langle L_P\rangle$ 
and the Dirac modes in QCD, and is 
mathematically valid on the temporally odd-number lattice 
in both confined and deconfined phases. 
From Eq.(\ref{eq:PLvsD}), we can investigate each Dirac-mode contribution 
to the Polyakov loop individually.

Remarkably, due to the factor $\lambda_n^{N_t -1}$, 
low-lying Dirac-mode contribution 
is negligibly small in RHS of Eq.(\ref{eq:PLvsD}),
compared to the other Dirac-mode contribution. 
In fact, the low-lying Dirac modes give little contribution 
to the Polyakov loop, regardless of confined or deconfined phase \cite{SDI13}.

Here, we give several comments on the relation (\ref{eq:PLvsD}) in order.
\begin{enumerate}
\item
Eq.(\ref{eq:PLvsD}) is manifestly gauge invariant,
because 
$
\langle n |\hat U_4|n\rangle =
\sum_s \langle n |s \rangle \langle s 
|\hat U_4| s+\hat t \rangle \langle s+\hat t|n\rangle
=\sum_s \psi_n^\dagger (s)U_4(s) \psi_n(s+\hat t)
$
is gauge invariant under the gauge transformation, 
$\psi_n(s)\rightarrow V(s) \psi_n(s)$.
\item
In RHS of Eq.(\ref{eq:PLvsD}), 
there is no cancellation between chiral-pair Dirac eigen-states, 
$|n \rangle$ and $\gamma_5|n \rangle$, because $(N_t-1)$ is even, i.e., 
$(-\lambda_n)^{N_t-1}=\lambda_n^{N_t-1}$, and  
$\langle n |\gamma_5 \hat U_4 \gamma_5|n\rangle
=\langle n |\hat U_4|n\rangle$. 
\item
The relation (\ref{eq:PLvsD}) is correct regardless of 
presence or absence of dynamical quarks, 
although dynamical quark effects appear in $\langle L_P\rangle$, 
the Dirac eigenvalue distribution $\rho(\lambda)$ and 
$\langle n |\hat U_4|n\rangle$.
\item
The relation (\ref{eq:PLvsD}) is correct also 
at finite density and temperature for any color number $N_c$.
\end{enumerate}

In lattice QCD simulations, 
we also numerically confirm Eq.(\ref{eq:PLvsD}) 
and quite small contribution of 
low-lying Dirac modes to the Polyakov loop 
in both confined and deconfined phases \cite{SDI13,DSI13}. 

From the analytical relation (\ref{eq:PLvsD}) and the numerical confirmation, 
we conclude that low-lying Dirac-modes 
give negligibly small contribution to the Polyakov loop, 
and are not essential for confinement, 
while these modes are essential for chiral symmetry breaking.
This conclusion indicates no direct one-to-one correspondence between 
confinement and chiral symmetry breaking in QCD.

\vspace{-0.15cm}
\section*{Acknowledgements}
\vspace{-0.15cm}
H.S. and T.I. are supported in part by the Grant for Scientific Research 
[(C) No.23540306, E01:21105006, No.21674002] 
from the Ministry of Education, Science and Technology of Japan. 

\vspace{-0.3cm}

\vspace{-0.15cm}

\begin{thebibliography}{99} 

\bibitem{NJL61} 
Y.~Nambu and G.~Jona-Lasinio, 
{\it Phys. Rev.} {\bf 122} (1961) 345; 
{\it Phys. Rev.} {\bf 124} (1961) 246.

\bibitem{SST95}
H.~Suganuma, S.~Sasaki and H.~Toki, 
{\it Nucl. Phys.} {\bf B435} (1995) 207; 
%%CITATION = HEP-PH/9312350;%%
{\it Prog. Theor. Phys.} {\bf 94} (1995) 373. 

\bibitem{M95W95}
O.~Miyamura, {\it Phys. Lett.} {\bf B353} (1995) 91;
R.M.~Woloshyn, {\it Phys. Rev.} {\bf D51} (1995) 6411.

\bibitem{GIS12}
S.~Gongyo, T.~Iritani and H.~Suganuma, 
{\it Phys. Rev.} {\bf D86} (2012) 034510;
T.~Iritani and H.~Suganuma, 
arXiv:1305.4049[hep-lat]; 
{\it PoS}~({\bf Lattice 2012}) (2012) 218; 
{\it PoS} ({\bf Confinement X}) (2013) 053.

\bibitem{LFKRS13}
P.M.Lo, B.Friman, O.Kaczmarek, K. Redlich, C. Sasaki, 
{\it Phys. Rev.} {\bf D88} (2013) 014506; {\it ibid.} 074502.

\bibitem{BC80}
T.~Banks and A.~Casher, {\it Nucl. Phys.} {\bf B169} (1980) 103. 

\bibitem{Rothe12} 
H.-J.~Rothe, {\it Lattice Gauge Theories}, 4th edition, 
World Scientific, 2012, and its references.

\bibitem{AFKS06}
Y. Aoki {\it et al.},
%Y. Aoki, Z. Fodor, S.D. Katz and K.K. Szabo, 
{\it Phys. Lett.} {\bf B643} (2006) 46; 
%Y. Aoki, G. Endrodi, Z. Fodor, S.D. Katz and K.K. Szabo, 
{\it Nature} {\bf 443} (2006) 675.

\bibitem{SDI13}
H.~Suganuma, T.M. Doi and  T.~Iritani, 
{\it PoS}~({\bf Lattice 2013}) (2013) 374;\\
{\it PoS}~({\bf QCD-TNT-III}) (2014) 042;
{\it Eur. Phys. J. Web of Conference} ({\bf ICNFP2013}) (2014).

\bibitem{DSI13}
T.M. Doi, H.~Suganuma, T.~Iritani, 
{\it PoS}~({\bf Lattice 2013}) (2013) 375; 
{\it PoS}~({\bf Hadron 2013}) (2013) 122.

\end{thebibliography}
\end{document}